\documentclass[preprint]{aastex}
\usepackage{emulateapj5}
\usepackage{apjfonts}

\usepackage{epsf}

\slugcomment{draft of \today}

\begin{document}

\def\etal{{\it et al.~}}
\def\eg{{\it e.g.,~}}
\def\ie{{\it i.e.,}}

\title{The Magnetohydrodynamic Kelvin-Helmholtz Instability:\\
A Three-Dimensional Study of Nonlinear Evolution\altaffilmark{4}}
 
\author{Dongsu Ryu\altaffilmark{1},
        T. W. Jones\altaffilmark{2},
        and Adam Frank\altaffilmark{3}}

\altaffiltext{1}
{Department of Astronomy \& Space Science, Chungnam National University,
Daejeon 305-764, Korea: ryu@canopus.chungnam.ac.kr}
\altaffiltext{2}
{Department of Astronomy, University of Minnesota, Minneapolis, MN 55455:
twj@msi.umn.edu}
\altaffiltext{3}
{Department of Physics and Astronomy, University of Rochester, Rochester
NY 14627: afrank@alethea.pas.rochester.edu}
\altaffiltext{4}
{The Astrophysical Journal, in press}

\begin{abstract}

We investigate through high resolution 3D simulations
the nonlinear evolution of compressible magnetohydrodynamic flows
subject to the Kelvin-Helmholtz instability. As in our earlier work
we have considered periodic sections of flows that contain a thin,
trans-sonic shear layer,
but are otherwise uniform. The initially uniform magnetic field is
parallel to the shear plane, but oblique to the flow itself.
We confirm in 3D flows the conclusion from our 2D
work that even apparently weak magnetic fields
embedded in Kelvin-Helmholtz unstable plasma flows can be fundamentally
important to nonlinear evolution of the instability. In fact, that statement
is strengthened in 3D by this work, because it shows how field line bundles
can be {\it stretched and twisted} in 3D as the quasi-2D {\it Cat's Eye}
vortex forms out of the hydrodynamical motions.
In our simulations twisting of the field may
increase the maximum field strength by more than a factor
of two over the 2D effect. If, by these developments, the Alfv\'en Mach
number of flows around the Cat's Eye drops to unity or less, our simulations
suggest magnetic stresses will eventually destroy the Cat's Eye and
cause the plasma flow to
self-organize into a relatively smooth and apparently stable
flow that retains memory of the original shear. For our 
flow configurations the regime in 3D for such reorganization is 
$4\lesssim M_{Ax} \lesssim 50$, expressed in terms of the Alfv\'en Mach number 
of the original velocity transition and the initial Alfv\'en speed projected to
the flow plan. When the initial field is stronger than this, either the flow
is linearly stable (if $M_{Ax} \lesssim 2$), or becomes stabilized
by enhanced magnetic tension due to the corrugated field along the shear layer
before the Cat's Eye forms (if $M_{Ax} \gtrsim 2$). For weaker fields
the instability remains essentially hydrodynamic in early stages,
and the Cat's Eye is destroyed by the hydrodynamic secondary instabilities
of a 3D nature. Then, the flows evolve into chaotic structures that  
approach decaying isotropic turbulence.
In this stage, there is considerable enhancement
to the magnetic energy due to stretching, twisting, and turbulent
amplification, which is retained long afterwards.
The magnetic energy eventually catches up to the kinetic
energy, and the nature of flows become magnetohydrodynamic.
Decay of the magnetohydrodynamic turbulence is enhanced
by dissipation accompanying magnetic reconnection.
Hence, in 3D as in 2D, very weak fields
do not modify substantially the character of the flow evolution, but 
do increase global dissipation rates.

\end{abstract}



\section{INTRODUCTION}

Strongly sheared boundary flows are ubiquitous in astrophysical
environments as diverse as the earth's magnetopause and supersonic jets.
The susceptibility of such boundaries to the Kelvin-Helmholtz (K-H)
instability is well-known (\eg Chandrasekhar 1961).
Development of the instability may lead to turbulence, momentum and energy
transport, dissipation and mixing of fluids (see, \eg Maslowe 1985 for a 
review).

Most astrophysical environments are electrically conducting,
so relevant fluids are likely to be magnetized on length and time scales 
of common interest. 
Thus, it is important to understand the role of magnetic fields in 
the K-H instability.
The basic linear stability analysis of the magnetohydrodynamic (MHD) K-H
instability was carried out long ago (\eg Chandrasekhar 1961; 
Miura \& Prichett 1982).
There is now also a growing literature of the nonlinear evolution of the MHD
K-H instability beginning from a variety of possible initial flow 
configurations, at least in the earlier evolution stages
in two dimensions (2D) 
(\eg Tajima \& Leboeuf 1980; Wang \& Robertson 1984; Miura 1984, 1987, 1997; 
Wu 1986; Dahlburg \etal 1997; Keppens \etal 1999; Keller \& Lysak 1999).
Fully three dimensional (3D) nonlinear studies are still quite limited
and so far have generally
not followed flow evolution to anything resembling a final state. They do
show that full coupling of magnetic field and flow in the third dimension
may quickly introduce obvious dynamical effects, however (\eg Galinsky 
\& Sonnerup 1994; Keppens \& T\'oth 1999; Keller, Lysak \& Song 1999).
The 3D simulations reported below, on the other hand, were
continued over many dynamical time scales, so that the ultimate relaxed
states for the flows are clear.

Strong magnetic fields, through their tension, are well known to
stabilize the K-H instability. However, the considerable potential for much
weaker fields to modify the nonlinear instability, and, in particular,
to reorganize the subsequent flow, has only recently been emphasized. 
Malagoli \etal (1996), Frank \etal (1996), Jones \etal (1997) and
Jeong \etal (2000) have carried out high resolution 2D and
$2\frac{1}{2}$D numerical MHD simulations of the full nonlinear
evolution of the K-H instability for periodic sections of 2D flows. They
have demonstrated clearly that an initially weak magnetic field can 
fundamentally alter evolution of the K-H instability, either by
disrupting the 2D hydrodynamical (HD) flow character or by enhancing
dissipation during the nonlinear evolution of this instability.

While specific issues may surely depend on matching details of a simulated
configuration to the physical situation imagined, the most basic 
insights often come from very simple, idealized model flows. In our
review here we follow Frank \etal (1996) and Jones \etal (1997)
in considering periodic sections of flows. Let us define the 2D
computational plane as the $x-y$ plane, so that for now there is an assumed
invariance along the $z$ direction. Initially, two uniform,
but oppositely directed, velocity fields along the $x-$direction are
separated by a thin, smooth trans-sonic shear layer ($M_0 = U_0/c_s = 1$, 
with $U_0$ the velocity difference across the shear layer). 
The magnetic field, which is initially
aligned within the $x-z$ plane, is uniform, and so is density. 
The most important
parameter in predicting the outcome of the subsequent MHD K-H instability
is the following Alfv\'enic Mach number of the velocity
transition; namely, $M_{Ax} = U_0/c_{Ax}$, with 
$c_{Ax} = B_x/\sqrt{\rho}$ the
projected Alfv\'en speed, and $B_x$ the magnetic field component
aligned with the flow in units giving magnetic pressure, $p_b = B^2/2$.
We note that Jones \etal (1997) demonstrated when $M_{Ax} \gtrsim 2$
(for which the instability is not stabilized by magnetic tension) and
the initial magnetic field is uniform that the existence of a finite
$B_z$ is largely irrelevant to evolution of the K-H instability;
that is, the orientation of the field in the $x-z$ plane does not,
by itself, matter, except through its isotropic pressure.

When $M_{Ax}\lesssim4$ but with the initial magnetic field aligned
with the flow, the field is a little too weak to provide linear
stability ($M_{Ax} \gtrsim 2$).
Modest growth of corrugations along the perturbed shear layer generates
sufficient magnetic tension to prevent further development of nonlinear
evolution. That is, the flow is {\it nonlinearly stabilized}. However, when
$M_{Ax} \gtrsim 4$, the magnetic field is too weak to have much, if any,
apparent influence during the linear and early 2D nonlinear phases of
the K-H instability. Thus, the initial development is largely HD.
So, a  {\it Kelvin's Cat's Eye} vortex forms with its axis
in the shear layer, but perpendicular to the initial flow. In 2D HD,
this structure is stable. Jones \etal (1997), therefore,
chose $M_{Ax}\sim4$ as a convenient boundary between {\it strong} and
{\it weak} magnetic field behaviors in the 2D MHD K-H instability.

Within the weak field regime, it is also possible to distinguish
further two qualitatively different evolutions.
Unless an initially uniform field is amplified sufficiently during one
rotation of the Cat's Eye to reduce $M_{Ax}$ to values of order unity
along the vortex perimeter, magnetic stresses have little immediate
dynamical influence on vortex evolution. Then, the magnetic field
primarily serves to enhance dissipation of kinetic energy through
expulsion of magnetic flux in the $x-y$ plane (a.k.a. flux annihilation,
via tearing mode reconnection). This case of {\it very weak field} was
called {\it dissipative} in Jones \etal (1997). In the discussion below
we will label flows initiated in this regime by {\bf VWF}. The more interesting
regime is that where the initial field is too weak to prevent formation
of the Cat's Eye, but strong enough that $M_{Ax} \sim 1$ at some
locations within the Cat's Eye by the end of a single vortex rotation.
Under those circumstances, relaxation of magnetic stresses during
reconnection deforms and then disrupts the Cat's Eye. This was called
the {\it weak field} regime or the {\it disruptive} regime in Jones
\etal (1997). We will label these cases below as {\bf WF}.

In a 2D flow, a magnetic field line is
stretched by about an order of magnitude while becoming wrapped around
a vortex that it once spanned. That reduces $M_{Ax}$ by a similar factor,
since the field strength increases proportionally to the length of a 
flux tube (\eg Gregori \etal 2000).
Thus, it turned out in 2D $M_{Ax}\sim20$ is the boundary between
the two cases. However, even for $M_{Ax} > 20$, there can be gradual
disruption of the Cat's Eye at late epoch by an accumulation of small
effects from Maxwell stresses. So, this dividing line is not distinct.

In the disruptive, {\bf WF} case, there is also a dynamical alignment
between the magnetic and velocity fields during reconnection along the
perimeter of the Cat's Eye, and local cross helicity ($|v\cdot B|$) is
maximized. In this configuration the 2D flow returns to a laminar form,
but is now stable to perturbations smaller than
the size of the computational box. 
Jones \etal (1997) emphasized that 2D vortex disruption was magnetically
driven, despite the fact that the initial $\beta = p_g/p_b >> 1$,
where $p_g$ is the thermal gas pressure. We note this, since it is very
common to ignore dynamical influences from magnetic fields under the
condition $\beta = p_g/p_b >> 1$. That measures only the relative
influences of pressure gradients, not the full Maxwell stresses.
The Alfv\'en Mach number, on the other hand, compares more closely
Maxwell to Reynolds stresses, so should provide a more direct measure
of the immediate dynamical consequences of the magnetic field in
nonequilibrium flows. Certainly, that is the case here.

Our objective now is to extend those previous 2D
results into fully 3D flows. This step is important, since
it is already well established that the Cat's Eye structure
so prominent and stable in plane-symmetric, 2D flows coming from the
HD K-H instability is unstable to perturbations
along its axis in 3D (Hussain 1984; Bayly 1986; Craik \& Criminale 1986).
The resultant HD flow becomes turbulent
(\eg Maslowe 1985). So, we should ask how a
weak magnetic field will modify that outcome. In addition, since
both vorticity and magnetic flux is subject to stretching in 3D
but not in 2D, and, since vortex stretching profoundly changes 3D flows
when compared to those in 2D, we might expect to find that as soon as the flow
evolution deviates from 2D character the characterizations listed
above no longer apply. We will find, in fact, that they do still apply,
but the domain of initial magnetic field strengths that can
significantly influence the flow evolution is extended to weaker
fields in 3D. We will also see that the morphologies and statistical
properties of magnetic and flow
structures expected during 3D nonlinear flow evolution depend on
the strength of the initial magnetic field. 
A preliminary report on some of these calculations is contained in
Jones \etal (1999), which also includes some useful animations
on a CD ROM. Those same animations are presently posted on the
web site: http://www.msi.umn.edu/$\sim$twj/research/mhdkh3d/nap98.html.
The plan of the present paper is as follows:
In \S 2 we will summarize the problem set-up and numerical method.
\S 3 contains detailed discussions of results.
A brief summary and conclusion follow in \S4.

\section{THE PROBLEM}

The simulations reported here are direct extensions of the 
Jones \etal (1997) study to fully 3D flows. The equations we solve
numerically are those of ideal compressible MHD, where the displacement
current and the separation between ions and electrons are neglected
as well as the effects of viscosity, electrical resistivity and thermal 
conductivity. In conservative form, the equations are
\begin{equation}
\frac{\partial \rho}{\partial t}+{\vec\nabla}\cdot\left(\rho{\vec v}\right)
= 0,
\end{equation}
\begin{equation}
\frac{\partial(\rho{\vec v})}{\partial t} + \nabla_j\cdot\left(\rho
{\vec v}v_j-{\vec B}B_j\right)+\nabla\left(p+{1\over2}B^2\right) = 0,
\end{equation}
\begin{equation}
\frac{\partial E}{\partial t} + {\vec\nabla}\cdot\left[\left(E+p+{1\over2}B^2
\right){\vec v}-\left({\vec v}\cdot{\vec B}\right){\vec B}\right] = 0,
\end{equation}
\begin{equation}
\frac{\partial \vec B}{\partial t} + \nabla_j\cdot\left({\vec B}v_j
-{\vec v}B_j\right) = 0,
\end{equation}
along with the constraint ${\vec\nabla}\cdot{\vec B}=0$ imposed to
account for the absence of magnetic monopoles (\eg Priest 1984).
Gas gas pressure is given by
\begin{equation}
p=\left(\gamma-1\right)\left(E-{1\over2}\rho v^2-{1\over2}B^2\right)_.
\end{equation}
Standard symbols are used for common quantities. The magnetic
pressure is $p_b = B^2/2$ and the Alfv\'en speed is $c_A =
B/\sqrt{\rho}$.

The simulations have been carried out in a cubic computational box
of length $L_x = L_y = L_z = L = 1$. Boundaries are periodic in the
directions contained within the shear layer (namely, $x$ and $z$)
and reflecting above and below the shear layer (namely, $y$). As
before we have simulated flows that are initially uniform except
for a hyperbolic tangent velocity shear layer in the $y$-coordinate,
given as
\begin{equation}
{{\vec v}_0} = - {{U_0}\over 2} \tanh\left({{y-{L_y/2}}\over a}\right)
{\hat x}
\end{equation}
with $a=L/25$. The equilibrium flow is directed in the $-x$
direction for $y > 0.5$, and $+x$ direction for $y < 0.5$. The velocity
difference across the shear layer is unity, $U_0 = 1$. The sonic Mach number
of the transition is unity, $M_s = 1$, and the adiabatic index, $\gamma = 5/3$.
With this configuration K-H unstable modes will have zero phase
velocities in the computational reference frame.
Whereas an initial $B_z$ has no appreciable
influence on $2\frac{1}{2}$D flows, because field lines could not
be stretched in that dimension, we expect field line stretching in
this dimension to be important in 3D. Thus, the initial magnetic field
is oblique to the flow direction, with $\theta=30\arcdeg$, but
parallel to the shear plane with strengths corresponding to
$M_{Ax} = 2.5,~5,~14.3,~50,~143,~500,$ and $1.43\times 10^3$.
For comparison, $\beta=(2/\gamma)(M_{Ax}\cos\theta\arcdeg/M_s)^2$ here.
See Table 1 for further details. A random perturbation of small
amplitude has been added to the velocity to initiate the instability.

All cases have been simulated with grids having $64^3$ and $128^3$ zones
(labeled `$l$' standing for `low' and `m' standing for `medium',
respectively in Table 1) to explore basic properties including
resolution issues. For three representative cases with
$M_{Ax} =  14.3,~50,~1.43\times 10^3$, the calculations have been repeated
again with $256^3$ zones (labeled `h' standing for `high' in Table 1). 

Each simulation has been run up to time $t = 20 \sim 50$ (see Figs.
3 and 8 for the end time).
For comparison, the sound crossing time  for the box is unity. 
With the initial
perturbation applied, the Cat's Eye forms by about $t \approx 6$ in
those cases where it develops. The nominal subsequent turnover time
for the Cat's Eye is also $t = t_e \sim 6$. We note for reference that our
computing time units here are 2.51 longer than those in Frank
\etal (1996) and Jones \etal (1997), since $L=2.51$ was set to there to
match earlier papers,
but they are same as those in Jeong \etal (2000). To aid comparison
we mention that in our present units the growth times associated
with the modes having wavelength $\lambda=L$, the box size, would
typically be $t_g \sim  0.6 - 0.7$. Thus, our simulations extend
typically $\sim 35 - 70$ K-H linear growth times for such modes.
Here, we have applied random velocity perturbations to the initial
equilibrium in the present simulations. 
Hence, modes with shorter wavelengths develop first if unstable. However,
they generally merge on time scales comparable to the growth time of the
mode with wavelength which is the sum of the wavelengths of merged modes.
So the mentioned growth time
is still a reasonable estimate of the time required for
instabilities to become a significant influence on the flow.

The ideal MHD equations have been solved using a multi-dimensional
MHD code based on the explicit, finite difference ``Total Variation
Diminishing'' or ``TVD'' scheme.  That method is an MHD extension of
the second-order finite-difference, upwinded, conservative gasdynamics
scheme of Harten (1983), as described by Ryu \& Jones (1995).
The multi-dimensional version of the code, along with a description of
various one and 2D flow tests is contained in Ryu, Jones, \& Frank (1995).
This version of the code contains a fast Fourier transform-based
``flux cleaning'' routine that maintains the ${\vec\nabla}\cdot{\vec B}=0$
condition at each time step within machine accuracy.

\section{RESULTS}

The seven cases listed in Table 1 include examples that in 2D exhibit
``dissipative'', ``disruptive'' and ``nonlinear stabilizing'' behaviors,
using the terminology defined in \S 1. Cases that are linearly stable were not
considered, since we expect no new behaviors. Cases 1-4, with
$50 \le M_{Ax} \le 1.43\times10^3$, would in 2D have been {\it dissipative}
in character, Cases 5-6, with $5 \le M_{Ax} \le 14.3$, would have been
{\it disruptive}, while Case 7, with $M_{Ax} = 2.5$, would have been
{\it nonlinearly stabilizing}.

Initially all the cases evolve in ways consistent with the
2D description in \S 1 for the same field strength. For Case 7 that is
basically the end of the story, since the flow is nonlinearly stabilized
and remains laminar through the entire simulation. Just as in the 2D
simulation described by Frank \etal (1996) the final result is a
broadened shear layer that has been subjected to very minimal kinetic
energy dissipation.

In all the remaining cases a Cat's Eye develops and is subsequently
destroyed. The Cat's Eye is a 2D structure, so up to the point of its
formation all the flows are still quasi-2D. Recall that in 2D the
Cat's Eye remained stable in the very weak field, {\bf VWF} case, since Maxwell
stresses were not built up enough to disrupt it. But in 3D this structure
is HD unstable (see $\S$3.1). However, by and large, we still
find the flow characters of {\it dissipative} (for the {\bf WF} case) and
{\it disruptive} (for the {\bf VWF} case) carry
over into 3D flows, as described in detail in \S 3.1 and 3.2 below.
One interesting deviation is that the 3D magnetic field in Case 4, 
with $M_{Ax} = 50$, is
significantly more disruptive than it would be in 2D. 
Case 4  would have been defined as
{\it dissipative} in 2D, and it exhibits similar properties in our low and 
medium resolution 3D simulations.   On the other hand, in higher
resolution 2D simulations 
it showed some evidence for long-term disruptive magnetic field
influence through accumulated small flow distortions.
Those tendencies are much more consequential in 3D, so, we will describe the
in detail the
3D behaviors observed in that case in \S 3.3.

\subsection{Very Weak Field ({\bf VWF}) Cases: Turbulence}

Since they come close to HD behaviors, and thus offer a useful
benchmark, we begin with discussion of Cases 1, 2 and 3
($M_{Ax} = 1.43\times10^3,~500,~143$), which would all fall under
the {\bf VWF} or {\it dissipative} descriptions in 2D.
Fig. 1 shows at two times the spatial distributions of magnetic
field strength ($|B|$) and vorticity magnitude ($|\omega|$) for Case 1$\ell$
and Case 1m, while Fig. 2 shows at three times the same
information for the analogous high resolution, Case 1h.
One can see that the global behaviors are qualitatively consistent in all
three simulations. Quantitatively, the comparisons are quite similar
to what we described previously in Frank \etal (1996) and in Jones
\etal (1997). That is, as expected, simulations with higher resolution capture
finer structures.

In 3D HD flow the Cat's Eye violently breaks up within approximately one
eddy turnover, and the flow becomes highly disordered, with very little
evidence of the initial shear field; \ie~decaying isotropic
turbulence develops (\eg Maslowe 1985). For the {\bf VWF} MHD
cases we studied (Cases 1, 2, and 3) this behavior is also seen.
We can identify the root causes for the disruption of the Cat's Eye
as follows: First, Hussain (1984) pointed out the
importance of the growth of coherent vortex tubes that span the
Cat's Eye. These features, called {\it rib vortices} by Hussain,
are clearly present in the early snapshots of Figs. 1 and 2.
Hussain pointed out that the ribs are anchored in saddle points
within the flow at the ends of the Cat's Eye, so they are subjected
to rapid and intense vortex stretching. This leads to non-axial
stresses on the flows.
A second effect is fluid elements caught in the Cat's Eye
vortex move along non-circular, or roughly elliptical, paths,
so that they feel time varying shear forces. Such fluid elements are
known to be subject to the {\it elliptical instability}, when motion
perpendicular to the elliptical path is allowed in 3D
(Bayly 1986; Craik \& Criminale 1986). Together these effects
unstably distort the initially 2D character of the flow, so that
the Cat's Eye breaks up violently in less than a single turnover.
Fig. 2 shows that by $t = 6$ the interior of the Cat's Eye in
Case 1h is filled with a knotted tangle of thin vortex tubes.
They are the remnants of smaller size vortices, which are developed
and gone through the processes of disruption and merger
earlier. Fig. 2 also illustrates for Case 1h that between
$t = 6$ and $t = 8$ the Cat's Eye has already been severely distorted.
By $t = 20$ the entire flow pattern has broken down into an
apparently isotropic distribution of vortex tubes.

At this point it is useful to look closely at the simultaneous
evolution of the magnetic field. Fig. 2 shows us that the regions of
strong magnetic field generally match the regions of strong vorticity
(although there are vortex tubes which are the remnants of the
initial vorticity in the problem or the early activity inside
the Cat's Eye as described more in below, and so do not match with
strong magnetic field). Note that magnetic fields in this case are
essentially passive, and the flows are nearly ideal. It is well-known
that under these circumstances the magnetic field and vorticity obey
the same evolution equation (\eg Shu 1992). So the coincidence of
strong magnetic field regions with strong vorticity region is, to
a certain degree, expected. We emphasize, however, that the coincidence
does not necessarily imply that the two vector fields are aligned.
In fact, that is distinctly {\it not} the case along the rib vortices
{\it as they first form}. Rather, magnetic field lines passing
diagonally across the shear layer are initially stretched around
the forming Cat's Eye and become embedded within the flow fields of
the rib vortices. At first the fields in the rib vortices are simply 
stretched over the ribs, but,
over the course of the Cat's Eye formation, those field lines 
become twisted around the
ribs, like twisted-pair electrical wires. This effect enhances
significantly the amount of stretching those lines undergo compared to
their 2D analogs. Thus, the magnetic fields embedded in the
rib vortices are substantially stronger than other field lines merely
stretched around the Cat's Eye perimeter. That explains the coincidence of
strong magnetic field regions with strong vorticity region at $t = 20$
in Fig. 2. It also enhances the role of weak magnetic fields in
3D compared to 2D, as we shall see.

As magnetic field lines become twisted around rib vortices they soon
develop topologies unstable to reconnection (\eg Lysak \& Song 1990),
however. The reconnection preserves helicity (\eg Ruzmaikin \&
Akhmetiev 1994), although other topological field measures,
such as twist, writhe  and kink that contribute to helicity may
change (\eg Berger \& Field 1984; Bazdenkov \& Sato 1998).
The product by the time the Cat's Eye begins to be disrupted is
a set of twisted magnetic flux tubes around the perimeter of
the Cat's Eye that do align themselves with the original rib vortices;
\ie in the $x-y$ plane.

Simultaneously, the complex motions in the Cat's Eye interior, which
were mentioned above, lead to extensive magnetic reconnection  events
that produce a 3D
version of magnetic {\it flux expulsion} mentioned in \S 1 (Weiss 1966;
see Jones \etal 1997 for discussion of that process).
The product, when the Cat's Eye
begins to break up, is a region of very weak and tangled magnetic field
inside the Cat's Eye that on average 
trends in the $z$ direction. The total magnetic flux through the
full computational box is constant, of course, but this series of
events has separated the magnetic flux embedded in the Cat's Eye into
relatively strong flux tubes perpendicular to the axis of the Cat's
Eye and wrapped around it, plus largely disordered magnetic flux inside the
Cat's Eye with a mean field aligned with the axis of the Cat's Eye.
This dichotomy is retained in the magnetic field structures at the end of
end of Case 4h, in fact, as we will address that in \S 3.3.

The evolution of energy partitioning is illustrated for
the {\bf VWF} cases along with Case 4 in Fig. 3.
Keep in mind that because we use a periodic box in the $x$ and $z$ directions
and hard walls in the $y$ direction, the system is effectively closed
and total energy is conserved. It is, therefore, a necessarily
decaying dynamical system, since there is finite numerical dissipation.
There is an abrupt, but almost imperceptible ($< 1\%$) decrease
in kinetic energy within the flow around $t \sim 6$, caused directly by
formation of the Cat's Eye. Beginning with Cat's Eye disruption,
however, there is a steady, steep decay of this quantity. 
By the end of this simulation the kinetic energy has dropped
by about two orders of magnitude or more. This is in sharp contrast to
the analogous 2D version of this K-H derived flow, where after
formation of the Cat's Eye the kinetic energy is virtually
constant on these time scales. Turbulent decay in 3D is the 
reason for the difference, of course. Our result is, indeed,
consistent with studies of 3D decaying MHD and HD turbulence
(\eg Mac Low \etal 1998; Stone \etal 1998; Porter \etal 1994),
which also showed rapid dissipation of kinetic energy.

Three points are noticed from the kinetic energy plot of Fig. 3.
First, in the medium resolution simulations, the decay rate
increases with increasing initial magnetic field, from Case 1
to Case 4. This is the the result of enhanced dissipation
through reconnection in MHD turbulence. Hence, even an obviously
very weak magnetic field with $M_{Ax} \gtrsim 50$ (or
$\beta \gtrsim 2.25\times10^3$ in our setup) does play an
important role as an agency of enhancing dissipation. This character of
increasing dissipation was also observed in 2D, although the
dissipation there occurred through reconnection around the
stable Cat's Eye instead of reconnection driven by turbulent
motion (Jones \etal 1997).
Second, in Case 1 the decay is faster in the high resolution simulation
than in the medium resolution calculation. This is because higher resolution
allows a greater number of smaller scale structures to form and, so,
reconnection events are more frequent.
Again, a similar behavior was observed in the 2D {\bf VWF}
cases (Jones \etal 1997).
Finally, the evolution curve of kinetic energy in Case 4 is quite different
in the two simulations with different resolution. This happens
because in the high resolution simulation, the magnetic field is
amplified enough locally to play a more important dynamical role. The
details are described in \S 3.3.

The magnetic energy plot of Fig. 3 shows the following behavior.
Initially the magnetic energy increases at the expense of the 
kinetic energy, but stops increasing before energy equipartition is reached.
After that, the magnetic energy starts decreasing, but the rate of decline
is smaller than that of the kinetic energy. During this period,
the flow character is close to that of HD turbulence. But eventually,
progressing from  smaller scales to larger scales (see the
discussion on energy power spectrum below) the magnetic
energy catches up the kinetic energy, and the character of
MHD turbulence is fully established. Then, both energies decay
with the same rate. The turbulence developed by the K-H instability
in a closed system is a decaying, quasi-isotropic turbulence,
and the symmetry of this flow does not
support dynamo action. Hence, the magnetic energy must decay on
some time scale along with the kinetic energy, and both
energies should convert into the thermal energy. Over a very
long time dependent on the effective magnetic Reynolds number at
the dissipation scale (see below for more discussion), but much longer
than our simulation, the magnetic field in this closed system
should return to something resembling the initial configuration. 

In simulations of ideal MHD flows, resistivity, $\eta$, is
provided by numerical truncation and diffusion at the grid-cell level.
So it does not have a constant value, but depends on the size of
structures considered. In a numerical code based on a second-order scheme,
such as the TVD scheme, the effective numerical resistivity is inversely
proportional to the square of the scale, ${\ell}$, $\eta \propto {\ell}^{-2}$
(Ryu \etal 1995). As a result, the effective magnetic Reynolds number
is proportional to the square of the scale, $R_m \propto {\ell}^2$.
We can use the evolution of magnetic energy in our simulations,
in fact, to estimate heuristically the effective magnetic Reynolds
numbers as follows. The magnetic energy decay rate for non-ideal
decaying incompressible MHD turbulence (\eg Biskamp 1993) is just
\begin{equation}
{dE_m\over dt} = - \eta \int j^2 d^3x
= -\eta \int (\nabla\times B)^2 d^3x \sim -2\eta {E_m\over L^2_3}_,
\end{equation}
where $L_3$ is the thickness of current sheet. From this we can write
roughly that the magnetic field decay time is
\begin{equation}
t_{dm} \sim {L^2_3\over2\eta} \sim {R_m L_3\over2 v}_,
\end{equation}
where
\begin{equation}
\eta \sim {v L_{diss}\over R_m}
\end{equation}
and $L_{diss}\sim L_3$ are used since $L_{diss}$ represents the scale
on which energy dissipation by reconnection occurs, that is the typical
thickness of current sheets.
Here, $v$ represents the typical flow velocity across the current sheet.
For the Case 1h simulation using $256^3$ grid zones, for instance,
we estimate from Fig.3 that $t_{dm} \sim 20$ and
$\sqrt{\left <v^2\right>} \sim \sqrt{2 E_k/\rho} \sim 5\times10^{-2}$
at $30 \lesssim t \lesssim50$. So for an effective magnetic Reynolds 
number corresponding to the typical scale of current sheet thickness,
$L_3\sim 10^{-2}L$ (see Fig. 6 for an estimate of $L_3$),
we obtain $R_m \sim 200$. Note that the smallest values for the scale 
$L_3$ correspond to $2-3$ grid zones, so they are numerically
limited. The inertial range of turbulence in a simulation should
require $R_m \gtrsim 10^3$. Hence, applying the inverse square
effective dissipation behavior of our second-order scheme,
an inertial range is possible on scales greater than roughly $\sim 8$ 
zones. That is, in our simulations, turbulence can be approximately
represented on scales larger than $\sim 8$ zones.

Additional insights about the evolution of fluid and magnetic
field properties can be gleaned from the 3D power spectra of
the kinetic end magnetic energies, $E_k(k)$ and $E_m(k)$,
respectively, defined as follows.
The Fourier amplitude of the kinetic energy is calculated as
\begin{equation}
A_{k,j}({\vec k}) = {1\over L_x L_y L_z}\int H_x H_y H_z \sqrt{\rho} v_j
\exp\left[i(k_x x + k_y y + k_z z)\right] d^3x,
\end{equation}
and similarly the Fourier amplitude of the magnetic energy
is calculated as
\begin{equation}
A_{m,j}({\vec k}) = {1\over L_x L_y L_z}\int H_x H_y H_z B_j
\exp\left[i(k_x x + k_y y + k_z z)\right] d^3x,
\end{equation}
where $j\in\{x,y,z\}$. Here, $H_x$, $H_y$ and $H_z$ are the Hanning
window functions (Press \etal 1986), which are given as
\begin{equation}
H_x = {1\over2}\left[1-\cos\left(2\pi{x\over L_x}\right)\right],
\end{equation}
and similarly for $H_y$ and $H_z$. Windowing in $y$ is used
because the flow is not periodic in that direction. Then, it seems
desirable to window in $x$ and $z$, too, in order avoid artificial
anisotropies in Fourier space. Assuming isotropy  on the scales of
interest (see Fig. 5
and discussion in below), the energy power spectra are given as
\begin{equation}
E_k(k) = {L_x L_y L_z \over 2 (2\pi)^3 W}
\left(|A_{k,x}(k)|^2+|A_{k,y}(k)|^2+|A_{k,z}(k)|^2\right)k^2,
\end{equation}
and
\begin{equation}
E_m(k) = {L_x L_y L_z \over 2 (2\pi)^3 W}
\left(|A_{m,x}(k)|^2+|A_{m,y}(k)|^2+|A_{m,z}(k)|^2\right)k^2,
\end{equation}
where
\begin{equation}
W = {1 \over L_x L_y L_z} \int H_x^2 H_y^2 H_z^2 d^3x.
\end{equation}
Note that with the above definition
\begin{equation}
\int E_k(k) dk = {1 \over L_x L_y L_z W} \int H_x^2 H_y^2 H_z^2
{1\over2}\rho v^2 d^3x,
\end{equation}
and
\begin{equation}
\int E_m(k) dk = {1 \over L_x L_y L_z W} \int H_x^2 H_y^2 H_z^2
{1\over2}B^2 d^3x,
\end{equation}
that is, $E_k(k)$ and $E_m(k)$ are the kinetic and magnetic energies
per unit $k$, respectively.

Fig. 4 shows the above two energy power spectra, along with their sum,
$E_{k+m}(k) \equiv E_k(k)+E_m(k)$, for Case 1h at $t = 6,~8,~12,~20,~32$
and $50$. Also, for comparison, we include lines with $E \propto k^{-5/3}$
and $E \propto k^{-3}$, representing the canonical forms for inertial
range 3D and 2D isotropic HD turbulence, respectively (\eg Lesieur 1997).
The vertical dotted lines indicate the scales of $L/4$, one fourth of
the box size, and 8 zones ($\log{k} = 0.602$ and $\log{k} = 1.505$).
Structures with $\ell \sim L$ (more specifically,
$\ell \gtrsim L/4$ according to our tests) have been strongly affected by the
finite box size, while structures with $\ell \lesssim 8$ zones have been
severely dissipated by numerical diffusion (see above). So we may regard
the region only between the two vertical dotted lines as an approximately
inertial range.
We have seen in Fig. 2 that at the earlier times, $t = 6$ and $8$,
there is still considerable large-scale, non-isotropic organization to
the flows (\eg the Cat's Eye). But later, at $t = 20$, the flow looks to the
human eye as though it is isotropic turbulence.
Fig. 4 supports that impression.
In particular, if we examine $E_{k+m}(k)$ in the inertial range,
we see that the power-law slope starts with a value close to $-3$,
as expected from
the 2D flow character of the Cat's Eye. Then, over time, 
the $E_{k+m}(k)$ becomes
flatter, but the slope is still steeper than $-5/3$, until the flow
develops into something very close to decaying isotropic turbulence
by $t\sim20$. After that, the amplitude decays with time, but the form
remains relatively unchanged to the end of the simulation.

Another point to emphasize is that at the early epochs $E_k(k)$ dominates
$E_m(k)$ on all scales. This fact is consistent with our earlier
conclusion that in the {\bf VWF} cases, the flow is initially
essentially HD in character. But, as complex flow structures
develop, magnetic field is amplified by flux stretching.
By $t\sim20$, $E_m(k)$
has caught up $E_k(k)$ on small scales. By the end of the simulation,
$E_m(k)\sim E_k(k)$ over the most of inertial range except on the largest
scales. Hence, by this time the flow of Case 1h shows the character
of MHD turbulence.

The evolution to a quasi-isotropic flow character in the {\bf VWF}
case can be seen by looking at Fig. 5. This shows for Case 1h
at a sequence of times $\left<v_x(y)\right>_{x,z}$, which is the
average of $v_x$ over the $x-z$ plane.
$\int\left<v_x(y)\right>_{x,z}dy$ is always very close to 0, from the 
symmetry of the initial conditions, although, since the initial
perturbations were random, there is no exact symmetry in $y$
required. The shear in $v_x$, represented by 
$d\left<v_x(y)\right>_{x,z}/dy$,
keeps decreasing rapidly. By $t = 40$, not only
$d\left<v_x(y)\right>_{x,z}/dy \approx 0$,
but also $\left<v_x(y)\right>_{x,z} \approx 0$ for all $y$.
This indicates there is no residual shear left, and
the flow has become isotropic.

There are a number of quantitative ways to characterize the
structural evolution of the flows in MHD simulations. The following
quantities are particularly
simple and useful:
The mean magnetic curvature radius, $L_1$,
\begin{equation}
L_1 \equiv \sqrt{\frac{\left<B^4\right>}{\left<\left[\left({\vec B}\cdot
{\vec\nabla}\right){\vec B}\right]^2\right>}}_,
\end{equation}
the flow Taylor microscale, $L_2$,
\begin{equation}
L_2 \equiv \sqrt{\frac{\left<v^2\right>}{\left<\left({\vec \nabla}\times
{\vec v}\right)^2\right>}}_,
\end{equation}
the magnetic Taylor microscale, $L_3$,
\begin{equation}
L_3 \equiv \sqrt{\frac{\left<B^2\right>}{\left<\left({\vec \nabla}\times
{\vec B}\right)^2\right>}}_,
\end{equation}
and the magnetic intermittency $I$,
\begin{equation}
I \equiv \frac{\left<B^4\right>}{\left<B^2\right>^2}_,
\end{equation}
(\eg Lesieur 1997 for $L_2$; Ethan T. Vishniac, private communication 1999
for others). The first of these, $L_1$, measures how sharply the magnetic
field lines are bent. $L_2$ and $L_3$ measure the transverse dimensions,
or thicknesses, of vortex tubes and current sheets, respectively. 
$I$ measures spatial contrast in the magnetic field strength
distribution; \ie~$I >>1$ signifies the presence of magnetic 
voids and relatively intense flux tubes.

Fig. 6 shows the evolution of the above quantities for the three
high resolution simulations Case 1h, 4h and 5h, which are
{\bf VWF}, {\bf VWF/WF} (transitional) and {\bf WF} cases, respectively 
(see the next two subsections for discussions on Cases 4h and 5h).
Initially the magnetic field is uniform, so that $L_1$ and $L_3$
are infinite, while $I = 1$. The initial shear layer gives
$L_2 \approx \sqrt{3La/4} \approx 0.17$.
We note that for the boundary conditions used the mean vector magnetic
field, $\langle{\vec B}\rangle$, is exactly constant; \ie~there is no
dynamo action. So, any net increase in magnetic energy must also lead
to the increase in the magnetic intermittency, $I>1$.
The rest of this paragraph focuses on Case 1h, the {\bf VWF} case.
One can see from $L_1$, $L_3$ and $I$ that very quickly, on a time
scale $t \lesssim 1 - 2$, the magnetic field is drawn into thin
structures. Especially, reduction of $L_1$ signals formation
of highly bent or twisted field regions. At the same time $L_2$ is reduced,
because vortices of smaller scales form. The modest increase in
$L_2$ just before $t \sim 4$ is due to the merger of smaller scale
vortices, whose remnants are seen at $t = 6$ in Fig. 2. By $t \sim 6-8$,
when the Cat's Eye is formed and begins to break up, magnetic intermittency,
$I$, is already very large.
Field curvature, $L_1$, stays low, but shows a peak around $t \sim 8$.
This is because the field is wrapped into the
Cat's Eye. However, $L_1$ increases due to partial field relaxation
during reconnection just prior to the Cat's Eye break-up.
During Cat's Eye break-up the field becomes
twisted and tangled, so that $L_1$ is again reduced.
After the Cat's Eye breaks up and the memory of the initial shear is
gone in this case ($t \gtrsim 20$), there begins a gradual relaxation in
all of $L_{1,2,3}$; 
that is, the size of vortex tubes increases and the field becomes
less strongly curved, while the thickness of current sheets increases
and the magnetic intermittency decreases. The last of these
remains relatively steady for Case 1h, near $I \sim 2$, for $t
\gtrsim 30$, while the others slowly increase to the end of that
simulation. This behavior reflects the fact at late time that magnetic field is
gradually relaxed by straightening itself, but flux tubes
remain stable structures. These properties match the finding of the slow decay
of magnetic energy for Case 1h in Fig. 3.

\subsection{Weak Field ({\bf WF}) Cases:
Magnetic Reorganization of the Cat's Eye}

This subsection discusses Cases 5 and 6 ($M_{Ax} = 14.3,~5$), which in
2D were categorized as the {\bf WF} or {\it disruptive} cases.
The character of 3D flow and magnetic field evolution in these cases
is perhaps best illustrated in the morphologies of Fig. 7.
It shows the spatial distributions of magnetic field strength ($B$)
and vorticity magnitude ($\omega$) in the high resolution simulation
Case 5h at three epochs. Initially, the Cat's Eye forms in this
case, so the morphologies at $t=6$, although more sheet-like, carry
some resemblance with those of Case 1h in Fig. 2. But, in Case 1h
the analogous images at $t = 20$ showed a completely disordered
arrangement of magnetic and vortex tubes. Here those features are
clearly laid out in patterns aligned to the original flow. Furthermore,
a closer examination shows both the magnetic flux tubes and the
vortex tubes to have a remarkable, sheet-like morphology with their
minimum extent in the $y$ direction. Thus, they follow and resemble the
original shear layer itself. There is also a good correspondence
between strong vorticity regions and strong magnetic field regions, as
one might expect in the presence of self-organization.

An interesting point for this simulation is that 2D cuts
through fixed $z$'s resemble very
much the 2D simulations of {\bf WF} cases (see the images in Frank
\etal 1996 and Jones \etal 1997). This is the result of the sheet-like
morphology, and the indication that the flow and magnetic field evolution,
although all three dimensions are available to the flows, the behavior
is essentially 2D in character.
So the {\bf WF} cases in 3D evolve towards some degree of self-organized
shear just as in 2D; that is, the Maxwell stresses developed during
formation of the Cat's Eye noticeably reorganize the flow and lead to
significant alignments between magnetic and velocity fields.
As for the {\bf VWF} cases (Cases 1-3), the magnetic field itself
becomes organized during Cat's Eye development through the action of
rib vortices into relatively strong-field flux tubes parallel to the
original velocity field, separated from relatively weak fields
trending along the Cat's Eye axis. Subsequently, the reorganized
velocity field aligns with the stronger magnetic field and retains
a clear {\it memory} of the original velocity shear.
Again, that contrasts with the vector fields
in the weakest field {\bf VWF} or HD cases in
3D, which become essentially isotropic in nature, excepting that the
mean vector magnetic field must remain unchanged, due to the symmetry.

The above point is obvious in Fig. 8, which shows at a sequence of times
$\left<v_x(y)\right>_{x,z}$ for the {\bf WF} cases as well as
the nonlinearly stable case (Case 7). The bottom panel in Fig. 8 shows
at $t = 30$ the shear strength, $d\left<v_x(y)\right>_{x,z}/dy$,
in the original midplane of the shear layer. Two points are made
from the figure. First, not only in the nonlinearly stable case but
also in the {\bf WF} cases, there left is still a well defined
shear layer which is also laminar. This is the result of
reorganization. Second, the residual shear strength at this time
clearly scales with the initial magnetic field strength (or more
importantly with $B_{x0}$). In HD, linear shear is stable against
linear perturbations but unstable to 3D finite-amplitude
perturbations (Bayly \etal 1988). But the magnetic field has the
stabilizing effects, just as in the MHD K-H instability case.
Stronger field can stabilize flows with larger linear shear.
The linear correlation of the residual shear with the initial field
strength is the direct consequence.

Among other things the self-organization and associated laminarity
in the {\bf WF} cases
substantially slow the rate of kinetic energy dissipation, since it
reduces the energy transfer to small, dissipation scales. That point
is clearly made by comparing Fig. 3 and Fig. 9, which illustrate the
evolution of energy partitioning for the {\bf VWF}, {\bf VWF/WF}
(transitional) and {\bf WF} cases.  We make three points from Fig. 9.
First, there is less kinetic energy dissipation in the stronger field
Cases 6 than in Case 5, as expected from the above discussion on the
residual shear.
Second, in both {\bf WF} Cases 5-6, about half of the initial kinetic
energy is still present at $t = 20$, and the decay rate has reduced
significantly from what it was during the time of Cat's Eye disruption.
So, this flow pattern should continue for a moderately
long time, but not as long as we found in 2D, since 
small scale structures in the third dimension can still form and enhance
the dissipation. Third, a comparison of Cases 5h and 5m shows a
good match between them. This indicates that small scale
structures do not play a major role, although they do exist. 
At the same time, by this measure, we can state safely that the
simulations are reasonably well resolved in the {\bf WF} cases.

The overall behaviors of the mean magnetic curvature radius, $L_1$,
the flow Taylor microscale, $L_2$, the magnetic Taylor microscale, $L_3$,
and the magnetic intermittency $I$ in Case 5h are similar
as those in Case 1h, as seen in Fig. 6. Three differences are noticed.
First, the $L$'s remain small in Case 1h since there is little dynamical
self-organization, But in Case 5h, self-organization relaxes
the magnetic field as well as vortices. As a result, $L$'s increase
after the Cat's Eye starts to break apart. Second, the small peak in
$L_2$ around $t\sim4$ is missing in the {\bf WF} case.
This is because initially the formation of smaller scale
vortices inside the Cat's Eye is not allowed due to the
magnetic field, although weak. This agrees with the visual
impression that structures are absent inside the Cat's Eye at $t=6$
in Fig. 7. Finally, $I$ approaches unity after $t\gtrsim20$,
indicating the magnetic field has
returned, more or less, to the initial uniform configuration.

\subsection{Case 4: A Transitional {\bf VWF/WF} Case
with Eventual Reorganization}

Case 4 with $M_{Ax} = 50$ begins with a magnetic field too weak in
2D to have any immediate direct dynamical role, although through
an accumulation of small magnetic field-induced perturbations even 
the 2D version of this case eventually begins to be distorted.
Thus, in 2D we would have classified this as a dissipative case
with a ``footnote''. Here we will use the label {\bf VWF/WF}.
On the face of it, the 3D Case 4 looks during formation of the Cat's Eye
like the {\bf VWF} cases discussed earlier, resembling
the morphologies in Figs. 1 and 2 at $t=6$. There is even a briefly
chaotic flow pattern right after the Cat's Eye breaks up. However,
in the high resolution simulation, Case 4h, slowly, over time,
the flow begins to reorganize, so that by the end of the
simulation residual shear becomes dominant, while the magnetic
field has organized into one predominant flux tube parallel to the
flow. This behavior is clearly visible in Fig. 10.
So, effectively, this case behaves like the {\bf WF}, disruptive
cases discussed in the immediately preceding subsection. This
case provides an evidence that the range of dynamically influential
magnetic fields is greater in 3D than in 2D.

The causes of that difference can be seen clearly by a closer examination
of magnetic field evolution during formation of the Cat's Eye. The key is
evident in Fig. 11, which at $t = 8$ shows regions where the
Alfv\'en Mach number is less than unity in Case 4h.
This is just as the Cat's Eye
begins to fall apart. The regions with $M_A < 1$ are all coincident
with rib vortices that initially were HD in character. Now,
however, they are magnetically dominated.  Within the rib vortices at
this time the flows are sub-Alfv\'enic, with mostly $0.1 \lesssim M_A < 1$. 
An image showing the regions
with small $\beta = p_g/p_b$ would be almost identical in appearance to
Fig. 11, as well. The smallest values of $\beta \gtrsim 1$, so it is really
the tension force rather than the pressure force that is revealing
the magnetic field's role.

In 2D we would have expected $M_A$ to drop by about one order of
magnitude from its initial value, since the magnetic field lines
around the vortex perimeter are stretched
by about that much due to formation and rotation of the Cat's Eye. 
That is insufficient to produce the properties seen in Fig. 11 and
consistent with the observation of Jones \etal (1997) that a 2D
$M_{Ax} \sim 50$ flow would not lead to magnetic dominance.
In 3D, however, formation of the rib vortices provides a new mechanism
to enhanced field amplification, as mentioned earlier. In particular,
field lines become wrapped around the rib vortices, so that they
become twisted as well as stretched around the Cat's Eye. 
Fig. 11 shows this effect by
tracing field lines within one rib vortex. Those field lines are
clearly twisted around the structure, so that this feature is a
legitimate flux tube with sufficient magnetic tension to begin
a self-organization of the flow field. A close examination of
the magnetic field distribution at this time reveals strengths
about ten times amplified over the initial field as expected;
namely, \ie~$|B| \gtrsim 0.2$ along the entire length of each of the
flux tubes visible in Fig. 11. But, each tube also contains a
core down much of its length that has $|B| \gtrsim 0.4$, an additional
enhancement we attribute to twisting. The fraction of the flow
under magnetic control at this time is still small, so the influence
is not immediately obvious. It is, however, crucial to the eventual
character of the flow. Evidently, if field amplification in any
significant region is able to reduce the Alfv\'en Mach number to less
than unity before the Cat's Eye is HD disrupted, some
memory of the original shear will be retained and, through
self-organization, the magnetic and flow fields will align, and
the flow may be smoothed. The evolution of these features for
Cases 4h and 5h is clearly seen in animations of vorticity and
magnetic pressure published in the CD ROM along with Jones \etal (1999),
and currently posted at the web site given at the end of \S 1.

The ways in which the above physics impacts on energy evolution are
shown for this case in Fig. 3. We see that in the high resolution
simulation energy dissipation is intermediate between the quasi-HD Case 1h,
which became turbulent, and Case 5h (in Fig. 9),
which quickly developed into a smooth flow.
We have noted previously that in medium and low resolution simulations
Case 4 behaves as
a {\bf VWF} flow, since then numerical dissipation prohibits
enough amplification of magnetic field to allow it to dominate dynamics.
In addition, we can see that in Case 4h the rate of kinetic energy
decay drops significantly after $t \sim 25$. By that time the flow
has begun to organize strongly, and initially numerous magnetic flux tubes,
twisted by vortical motion, have merged into a single, relatively
intense structure. Note in this respect from Fig. 11 that the
magnetic energy, $E_m$, in Fig. 3 is relatively constant from that
time on, as well.

Fig. 12 shows images of the magnetic flux structures of Case 4h at
$t = 40$. The dominant flux tube is obvious. It contains most of
the original flux that passed through the $x = 0$ and $x = L_x$ faces
of the computational box. Originally all the field lines ran obliquely
in the $x-z$ plane, but now most of the magnetic energy is
concentrated in this one structure, aligned in the $x$ direction alone.
On the other hand, for the periodic boundary conditions applied here 
the magnetic flux through each of the individual faces of the
computational box is preserved. So, there must have been a
topological change in the magnetic field along the way.
That fact can also be seen in Fig. 12, where we see most of 
the flux through the $z = 0$ and $z = L_z$ faces 
is now provided by field lines that meander around this
dominant flux tube, and essentially orthogonal to it. 
That is, the magnetic flux has separated as a result of
the instability and self-organization into two distinct domains.

Examination of the various structure measures in Fig. 6 augments the
sense that Case 4h represents a transition between the quasi-HD
Case 1 and the immediately reorganized Case 5. The magnetic field 
curvature measure, $L_1$, and the magnetic Taylor microscale, $L_3$,
evolve in very similar ways for Cases 4 and 5, reflecting the fact
that the magnetic field in each case is strong enough to smooth the
flow and coalesce into one dynamically important flux tube. On the other
hand, there is a much closer match between Cases 1 and 4 with regard
to the flow Taylor microscale, $L_2$, reflecting the development of
chaotic flow in each of these cases. The magnetic intermittency, $I$,
stays moderately large, approaching $\sim 2-3$ by the end of the simulation,
confirming the formation of one big flux tube.

\section{SUMMARY AND DISCUSSION}

Through high resolution MHD simulations using up to $256^3$ grid zones,
we have studied the 3D nonlinear evolution of the compressible MHD K-H
instability. As in our earlier work, we have considered periodic
sections of flows that contain a thin shear layer, but are otherwise
uniform. The initially uniform magnetic field is parallel to the
shear plane, but $30\arcdeg$ oblique to the flow itself. Its strength
spans the range corresponding to $M_x=2.5\sim1.43\times 10^3$.
The sonic Mach number of the flow transition was initially unity.

The most important consequence of this work is confirmation in
3D flows of the conclusion from our earlier 2D work (Frank \etal 1995;
Jones \etal 1997; Jeong \etal 2000) that {\it even apparently weak
magnetic fields corresponding to $M_{Ax} \gtrsim 4$ can be
important to nonlinear evolution of the K-H instability}.
The role of weak magnetic fields has been manifested in the following
two ways. First, in the case of {\it very weak field} ({\bf VWF} or
{\it dissipative} case) with $M_{Ax} \gtrsim 50$, dissipation is
enhanced through magnetic reconnection. In this case, the instability
remains essentially HD in character to the end. That is,
the Cat's Eye is destroyed by HD secondary instabilities
and the flows are developed into mostly isotropic turbulence.
But the decay rate of the turbulence increases. Second, in the case
of {\it weak field} ({\bf WF} or {\it disruptive} case) with
$4\lesssim M_{Ax} \lesssim 50$, the Cat's Eye is destroyed by the
magnetic stress of the field which has been amplified by stretching
and twisting on the perimeter of the vortex, once the Alfv\'en Mach
number of flows around the Cat's Eye drops to unity or less.
The flows, in this case, are eventually self-organized into relatively
smooth ones with linear shear, which are stable against further
instabilities.

There are two noticeable differences in the results of the current 3D
work from that of our previous 2D work. First, 
in the {\bf VWF} cases, while the Cat's Eye remains stable in 2D,
it is destroyed in 3D by inherently 3D instabilities
(Hussain 1984; Bayly 1986; Craik \& Criminale 1986). This is a
HD process, which was fully studied before.
The second difference, a 3D MHD process,
is additional amplification of magnetic
field by {\it twisting} inside rib vortices developed hydrodynamically
around the Cat's Eye. That leads to an increased role
for magnetic stresses in destroying the Cat's Eye. 
In both 2D and 3D {\bf WF} cases
magnetic field caught in the initially quasi-HD
roll-up of the Cat's Eye vortex tubes within the initial shear layer
is amplified by field-line {\it stretching}. In 2D the field strength on
the perimeter of the vortex is increased by about an
order of magnitude, representing the increased length of field
lines dragged around the forming Cat's Eye, before they become subject to 
magnetic reconnection. In 3D that effect is further enhanced by the
development of rib vortices spanning the Cat's Eye that twist field
lines into flux tubes, which then span the Cat's Eye and apply a
tension force to the plasma. In our 3D simulation, twisting of the field
increases the maximum field strength by more than a factor
of two over the 2D effect. For our rather idealized uniform density
and vector field configurations, the boundary field strength for the {\bf WF}
case decreases to the value corresponding to $M_{Ax} \sim 50$ in 3D from
$M_{Ax} \sim 20$ in 2D.

Two additional interesting points can also be made.
First, in the {\bf WF} cases, the magnetic
energy reaches its maximum just as magnetic stresses begin to destroy
the Cat's Eye.
However, it returns very close to its initial value, becoming almost
uniform magnetic field again, on the time scale same as the flow becomes
organized. The latter is barely longer than
the time needed to form the Cat's Eye
and disrupt it. Hence, the field effectively plays the role of a
{\it catalyst}.
Second, in the {\bf VWF} cases, where the flows become turbulent
and, so, are not re-organized, the amplified magnetic field during the
development of turbulence is retained long afterwards. The magnetic
energy decays slowly, until it catches up to the kinetic energy.
Then the flows approach decaying MHD turbulence, 
so that the magnetic energy along with the
turbulent kinetic energy decays at an enhanced rate.

In the cases of strong field with $M_{Ax} \lesssim 4$, the development
of the MHD K-H instability is essentially 2D in character, even though
variation along the third dimension is allowed. When
$M_{Ax} \lesssim 2$, the MHD flow is linearly stable and the instability
is not initiated. When $2\lesssim M_{Ax} \lesssim 4$, the shear layer is
initially corrugated, but the enhanced magnetic tension due to the
corrugated magnetic field stabilizes the instability before the Cat's
Eye forms; that is, the flow is nonlinearly stable.

The model configurations studied have no global helicity,
and, thus, are not capable of dynamo action.
Indeed the mean vector field is a constant throughout the simulations.
So, the enhancement of magnetic energy comes from
twisting followed by stretching of magnetic field lines, and/or from
maintenance of significant magnetic intermittency (non-uniformity),
not through generation of a large-scale field.
We have seen that in the transitional {\bf VWF/WF} case, however,
the enhancement of magnetic energy is maintained
through concentration of magnetic flux in flux tubes.
Beyond the dynamical impact of such flux concentrations,
that tendency could also be
significant astrophysically for a different reason. In particular,
measures of magnetic field strength, such as Faraday rotation, Zeeman
splitting and the intensity of synchrotron emission become biased to those
localized structures, so it becomes important to establish the intermittency of
the field to understand associated observations.

We note that our simulations have been done in an idealized box with
periodic boundaries along the $x$ and $z$-directions and reflecting
boundaries along the $y$-direction.
As a result, they have the following practical limitations.
First, some astrophysical systems subject to the K-H instability,
such as jets, contain continuous supplies of kinetic energy, while
our simulations conserve the total energy.
This limitation may be overcome by considering the {\it convective}
K-H instability, which employs inflow/outflow boundaries along the
$x$-direction, as Wu (1986) did in his 2D simulations.
However, since one must then use a much larger computational domain
to contain the evolving structures, the simulations become
substantially more expensive, and it would be currently possible to 
follow only up to the
early stage of the nonlinear development of the instability.
A second constraint is due to the periodic nature of the $z$-boundaries,
along the shear plane but perpendicular to the flow direction. From
this symmetry the axis of the 
Cat's Eye is constrained to be perpendicular to the initial flow 
direction. If the Cat's Eye
were allowed to rotate relative to the background flow, it would interact
with the flow by crumpling or corrugating. This might lead somewhat
different initial nonlinear behaviors in the {\bf WF} and {\bf VWF/WF} cases.
Relaxation of that symmetry must await later work.
Finally, the reflecting boundaries along the $y$-direction imposes
another limitation of our simulations.
However, in 2D, we saw that when activities are limited around the
shear boundary as in the {\bf WF} case, the effects of the reflecting
boundaries are minimal (see, Frank \etal 1996; Malagoli \etal 1996).
In addition, in the {\bf VWF} case, flow develops into turbulence,
so we expect the boundary effects would not be very important to the 
local properties of the flow.

Thus, we encourage restraint in direct application of our results for 
interpretation of observational features. 
Our intent is rather to provide more general physical insights into 
boundary layer dynamical processes within astrophysical objects subject to
the K-H instability, such as jets associated with young stellar
objects, accreting binaries, or larger scale flows from active galaxies 
(\eg Ferrari \etal 1980), strongly sheared flows in the solar
corona (\eg Kopp 1992) and the earth's magnetopause separating
the magnetosphere from the solar wind (\eg Miura 1984).
In addition, our findings have several broader implications in astrophysics.
The most obvious is that relatively weak magnetic fields may be able to
reduce the development of turbulence from the K-H
instability and diminish the tendency for mixing and related kinds of
transport across slip surfaces. This work, thus, augments earlier 
suggestions that weak magnetic fields may inhibit turbulent diffusion 
(\eg Vainshtein \& Rosner 1991).

\acknowledgments

The work by DR was supported in part by KOSEF through
grant and 981-0203-011-2.
The work by TWJ was supported in part by the NSF through grants
INT95-11654 and AST96-19438, by NASA grant NAG5-5055 and by the
University of Minnesota Supercomputing Institute.
The work by AF was supported by NSF Grant AST-0978765 and NASA Grant
NAG5-8428.
We thank the anonymous referee for clarifying comments.



\begin{deluxetable}{clllll}
\footnotesize
\tablecolumns{7}
\tablecaption{Summary of Initial Configurations}
\tablehead{\colhead{{Case}\tablenotemark{a}} & \colhead{$B_x$\tablenotemark{b}}
& \colhead{$M_{Ax}$\tablenotemark{c}} & \colhead{$M_{A0}$\tablenotemark{d}}
& \colhead{$\beta$\tablenotemark{d}} 
& \colhead{$N_{\rm grid}$} }
\startdata
1 ({\bf VWF}) & 0.0007 & 1430 & $1.24\times10^3$ & $1.84\times10^6$
& $64^3$ (1$l$), $128^3$ (1m), $256^3$ (1h) \\
2 ({\bf VWF}) & 0.002  & $500$              & $433$            & $2.25\times10^5$
& $64^3$ (2$l$), $128^3$ (2m)               \\
3 ({\bf VWF}) & 0.007  & $143$              & $124$            & $1.84\times10^4$
& $64^3$ (3$l$), $128^3$ (3m)               \\
4 ({\bf VWF/WF}) & 0.02   & $50$               & $43.3$           & $2.25\times10^3$
& $64^3$ (4$l$), $128^3$ (4m), $256^3$ (4h) \\
5~ ({\bf WF}) & 0.07   & $14.3$             & $12.4$           & $184$
& $64^3$ (5$l$), $128^3$ (5m), $256^3$ (5h) \\
6 ~({\bf WF}) & 0.2    & $5$                & $4.33$           & $22.5$
& $64^3$ (6$l$), $128^3$ (6m)               \\
7 ~({\bf SF}) & 0.4    & $2.5$              & $2.17$           & $5.63$
& $64^3$ (7$l$), $128^3$ (7m)               \\
\enddata
\tablenotetext{a} {{\bf VWF}, {\bf WF} and {\bf SF} labels refer to
``very weak field'', ``weak field'' and ``strong field'' behaviors,
respectively, found {\it for analogous  2D simulations}, as defined in $\S$1.
All models used $c_s=1$, $M_s=U_0/c_s=1$, $L=1$, $a=L/25$,
$\rho_0 = 1$, and $\gamma=5/3$.}
\tablenotetext{b}{$B_y=0$ and $B_z=B_x\tan\theta$ with $30\arcdeg$
were used.}
\tablenotetext{c}{$M_{Ax} = U_0\sqrt{\rho_0}/B_x$.}
\tablenotetext{d}{$M_{A0}$ and $\beta$ are defined from the total initial
uniform magnetic field strength, \ie ~$B=\sqrt{B_x^2+B_z^2}$.}
\end{deluxetable}


\begin{figure}
\figcaption{Volume rendering of the strong magnetic field ($B$)
and vorticity magnitude ($\omega$) structures in the medium resolution
simulation Case 1m and in the low resolution simulation Case 1$l$
(very weak field, or {\bf VWF} case).
Darker regions correspond to higher values and the gray scale was set 
arbitrarily to highlight structures.}
\end{figure}

\begin{figure}
\figcaption{Volume rendering of strong magnetic field ($B$) and
vorticity magnitude ($\omega$) structures in the high resolution
simulation Case 1h (very weak field, or {\bf VWF} case) at several epochs.
Darker regions correspond to higher values and the gray scale was set 
arbitrarily to highlight structures.}
\end{figure}

\begin{figure}
\figcaption{Energy evolution in the high resolution simulations Case 1h
and 4h and in the medium resolution simulations Case 1m to 4m
({\bf VWF} cases and a transitional case).
Shown are the normalized thermal, kinetic, and magnetic energies.}
\end{figure}

\begin{figure}
\figcaption{Temporal evolution of energy spectra in the high resolution
simulation Case 1h ({\bf VWF} case).
Shown are the spectra of kinetic energy ($E_k$), magnetic energy ($E_m$),
and kinetic plus magnetic energy ($E_{k+m}$).
For comparison, solid lines draw $k^{-5/3}$ and $k^{-3}$ power laws.
See text for the definition of energy spectra.}
\end{figure}

\begin{figure}
\figcaption{Temporal evolution of the averaged shear velocity profile in
the high resolution simulation Case 1h ({\bf VWF} case).}
\end{figure}

\begin{figure}
\figcaption{Evolution of some global structure measures in three high
resolution simulations (Case 1h, 4h, 5h).
Shown are the magnetic curvature radius ($L_1$), the flow Taylor
microscale ($L_2$), the magnetic Taylor microscale ($L_3$), and the
magnetic intermittency ($I$).
See text for the definitions.}
\end{figure}

\begin{figure}
\figcaption{Volume renderings of strong magnetic field ($B$) and
vorticity magnitude ($\omega$) structures in the high resolution
simulation Case 5h ({\bf WF} case) at several epochs.
Darker regions correspond to higher values and the gray scale was set 
arbitrarily to highlight structures.}
\end{figure}

\begin{figure}
\figcaption{Temporal evolution of the averaged shear velocity profile in
the high resolution simulation Case 5h and in the medium resolution
simulations Case 5m to 7m ({\bf WF} cases and strong field case).
The bottom panel (e) shows the derivative of the averaged shear velocity
around $y=L/2$ at $t=30$ in Case 5h and 5m and at $t=20$ in Case 6m and 7m.}
\end{figure}

\begin{figure}
\figcaption{Energy evolution in the high resolution simulation Case 5h
and in the medium resolution simulations Case 5m and 6m ({\bf WF} cases).
Shown are the normalized thermal, kinetic, and magnetic energies.}
\end{figure}

\begin{figure}
\figcaption{Volume renderings of strong magnetic field ($B$) and
vorticity magnitude ($\omega$) structures in the high resolution
simulation Case 4h (transitional case) at several epochs.
Darker regions correspond to higher values and the gray scale was set arbitrary
to highlight structures.}
\end{figure}

\begin{figure}
\figcaption{Volume rendering showing regions at $t = 8$ for Case 4h
(transitional case), where the Alfv\'en Mach number is less than unity.
These regions also trace out rib vortices and are twisted magnetic flux
tubes. One such tube structure is identified by a bundle of magnetic
field lines are traced in gray.}
\end{figure}

\begin{figure}
\figcaption{Two iso-surfaces (semi-transparent) highlighting regions of
strong to moderate magnetic field strength along with several
selected magnetic field lines in the high resolution simulation Case 4h
(transitional case) at $t=40$. Note how one bundle of field lines
threads the axis of one of the magnetic strength isosurfaces, which is
aligned with the $x$ axis. The other field lines are all chosen to
originate on the $z = 0$ face. They meander, but on average are orthogonal
to the strong field structure.}
\end{figure}

\end{document}